\documentclass[aps,twocolumn,prl,superscriptaddress,showpacs]{revtex4}
\usepackage{epsfig, amssymb}

\begin{document}
\advance\hoffset by  -4mm

\newcommand{\de}{\Delta E}
\newcommand{\mbc}{M_{\rm bc}}
\newcommand{\bb}{B{\bar B}}
\newcommand{\qq}{q{\bar q}}
\newcommand{\ks}{\bar{K}^0}
\newcommand{\kshort}{K^0_S}
\newcommand{\kstar}{\bar{K}^{*0}}
\newcommand{\kz}{\bar{K}^{(*)0}}
\newcommand{\kpi}{K^-\pi^+}
\newcommand{\kpipin}{\kpi\pi^0}
\newcommand{\kpipipi}{\kpi\pi^-\pi^+}
\newcommand{\dkpi}{D^0\to\kpi}
\newcommand{\dkpipin}{D^0\to\kpipin}
\newcommand{\dkpipipi}{D^0\to\kpipipi}
\newcommand{\bdnks}{\bar{B}^0\to D^0 \ks}
\newcommand{\bdnkstar}{\bar{B}^0\to D^0 \kstar}
\newcommand{\bdsks}{\bar{B}^0\to D^{*0} \ks}
\newcommand{\bdskstar}{\bar{B}^0\to D^{*0} \kstar}
\newcommand{\bdnkz}{\bar{B}^0\to D^0 \kz}
\newcommand{\bdskz}{\bar{B}^0\to D^{*0} \kz}
\newcommand{\bdzks}{\bar{B}^0\to D^{(*)0} \ks}
\newcommand{\bdzkstar}{\bar{B}^0\to D^{(*)0} \kstar}
\newcommand{\bdzkz}{\bar{B}^0\to D^{(*)0} \kz}
\newcommand{\bdbarkstar}{\bar{B}^0\to\bar{D}^0 \kstar}
\newcommand{\bdsbarkstar}{\bar{B}^0\to\bar{D}^{*0} \kstar}
\newcommand{\bdzbarkstar}{\bar{B}^0\to\bar{D}^{(*)0} \kstar}
\newcommand{\dsdpi}{D^{*0}\to D^0\pi^0}
\newcommand{\bdppi}{\bar{B}^0\to D^+\pi^-}
\newcommand{\br}{{\cal B}}

\title{\Large \rm 
Observation of $\bdnks$ and $\bdnkstar$ decays}

\affiliation{Budker Institute of Nuclear Physics, Novosibirsk}
\affiliation{Chiba University, Chiba}
\affiliation{Chuo University, Tokyo}
\affiliation{University of Cincinnati, Cincinnati, Ohio 45221}
\affiliation{University of Frankfurt, Frankfurt}
\affiliation{Gyeongsang National University, Chinju}
\affiliation{University of Hawaii, Honolulu, Hawaii 96822}
\affiliation{High Energy Accelerator Research Organization (KEK), Tsukuba}
\affiliation{Hiroshima Institute of Technology, Hiroshima}
\affiliation{Institute of High Energy Physics, Chinese Academy of Sciences, Beijing}
\affiliation{Institute of High Energy Physics, Vienna}
\affiliation{Institute for Theoretical and Experimental Physics, Moscow}
\affiliation{J. Stefan Institute, Ljubljana}
\affiliation{Kanagawa University, Yokohama}
\affiliation{Korea University, Seoul}
\affiliation{Kyoto University, Kyoto}
\affiliation{Kyungpook National University, Taegu}
\affiliation{Institut de Physique des Hautes \'Energies, Universit\'e de Lausanne, Lausanne}
\affiliation{University of Ljubljana, Ljubljana}
\affiliation{University of Maribor, Maribor}
\affiliation{University of Melbourne, Victoria}
\affiliation{Nagoya University, Nagoya}
\affiliation{Nara Women's University, Nara}
\affiliation{National Lien-Ho Institute of Technology, Miao Li}
\affiliation{National Taiwan University, Taipei}
\affiliation{H. Niewodniczanski Institute of Nuclear Physics, Krakow}
\affiliation{Nihon Dental College, Niigata}
\affiliation{Niigata University, Niigata}
\affiliation{Osaka City University, Osaka}
\affiliation{Osaka University, Osaka}
\affiliation{Panjab University, Chandigarh}
\affiliation{Peking University, Beijing}
\affiliation{RIKEN BNL Research Center, Upton, New York 11973}
\affiliation{Saga University, Saga}
\affiliation{University of Science and Technology of China, Hefei}
\affiliation{Seoul National University, Seoul}
\affiliation{Sungkyunkwan University, Suwon}
\affiliation{University of Sydney, Sydney NSW}
\affiliation{Tata Institute of Fundamental Research, Bombay}
\affiliation{Toho University, Funabashi}
\affiliation{Tohoku Gakuin University, Tagajo}
\affiliation{Tohoku University, Sendai}
\affiliation{University of Tokyo, Tokyo}
\affiliation{Tokyo Institute of Technology, Tokyo}
\affiliation{Tokyo Metropolitan University, Tokyo}
\affiliation{Tokyo University of Agriculture and Technology, Tokyo}
\affiliation{Toyama National College of Maritime Technology, Toyama}
\affiliation{University of Tsukuba, Tsukuba}
\affiliation{Utkal University, Bhubaneswer}
\affiliation{Virginia Polytechnic Institute and State University, Blacksburg, Virginia 24061}
\affiliation{Yokkaichi University, Yokkaichi}
\affiliation{Yonsei University, Seoul}
  \author{P.~Krokovny}\affiliation{Budker Institute of Nuclear Physics, Novosibirsk} 
  \author{K.~Abe}\affiliation{High Energy Accelerator Research Organization (KEK), Tsukuba} 
  \author{T.~Abe}\affiliation{Tohoku University, Sendai} 
  \author{I.~Adachi}\affiliation{High Energy Accelerator Research Organization (KEK), Tsukuba} 
  \author{H.~Aihara}\affiliation{University of Tokyo, Tokyo} 
  \author{M.~Akatsu}\affiliation{Nagoya University, Nagoya} 
  \author{Y.~Asano}\affiliation{University of Tsukuba, Tsukuba} 
  \author{T.~Aso}\affiliation{Toyama National College of Maritime Technology, Toyama} 
  \author{V.~Aulchenko}\affiliation{Budker Institute of Nuclear Physics, Novosibirsk} 
  \author{T.~Aushev}\affiliation{Institute for Theoretical and Experimental Physics, Moscow} 
% ...
\collaboration{The Belle Collaboration}

\begin{abstract}
We report on a search for $\bdzkz$ decays based on
$85\times 10^6$ $\bb$ events
collected with the Belle detector at KEKB.
The $\bdnks$ and $\bdnkstar$ decays have been observed for the first
time with the branching fractions
$\br (\bdnks)    =(5.0^{+1.3}_{-1.2}\pm 0.6)\times 10^{-5}$ and
$\br (\bdnkstar) =(4.8^{+1.1}_{-1.0}\pm 0.5)\times 10^{-5}$.
No significant signal has been found for the $\bdskz$ and $\bdzbarkstar$ 
decay modes, and upper limits at 90\% CL are presented.
\end{abstract}
\maketitle

\clearpage
{\ }
\clearpage

{\renewcommand{\thefootnote}{\fnsymbol{footnote}}}
\setcounter{footnote}{0}

Since the recent discovery of CP violation in the $B$ meson system,
through the measurement of non-zero values for 
$\sin 2\phi_1$~\cite{sin2phi}, attention
has turned towards the measurement of the other Unitary 
Triangle angles. Such measurements will allow tests of the 
Kobayashi-Maskawa ansatz and of the Standard Model. 
Precise measurements of the branching fraction for $\bdnkstar$, 
$\bdbarkstar$ and ${\bar B^0}\to D^0_{CP} {\bar{K}^{*0}}$ decays, 
where $D^0_{CP}$ denotes $D^0$ or $\bar{D}^0$ decay to a CP eigenstate, 
will allow a measurement of the angle $\phi_3$~\cite{dunietz}.
The decay $\bdnks$ can also be used to measure time-dependent
CP asymmetry in $B$ decays~\cite{gronau}.
So far no experimental information is available for any of these decays.

In this Letter we report on a search for the $\bdzks$, $\bdzkstar$ and
$\bar{B}^0\to \bar{D}^{(*)0}\kstar$~\cite{conj}
decays with the Belle detector~\cite{NIM} at the KEKB asymmetric energy 
$e^+e^-$ collider~\cite{KEKB}. The results are based on a 78~fb$^{-1}$
data sample collected at the center-of-mass (CM) energy of the 
$\Upsilon(4S)$ resonance, which  contains $85\times 10^6$ produced 
$\bb$ pairs. 


The Belle detector has been described elsewhere~\cite{NIM}.
Charged tracks are selected with a set of requirements based on the
average hit residual and impact parameter relative to the
interaction point (IP). We also require that the transverse momentum of
the tracks be greater than 0.1 GeV$/c$ in order to reduce the low 
momentum combinatorial background.

For charged particle identification (PID), the combined information
from specific ionization in the central drift chamber ($dE/dx$),
time-of-flight scintillation counters (TOF) and aerogel \v{C}erenkov
counters (ACC) is used.
At large momenta ($>2.5$~GeV$/c$) only the ACC and $dE/dx$ are used.
Charged kaons are selected with PID criteria that have
an efficiency of 88\%, a pion misidentification probability of 8\%,
and negligible contamination from protons.
All charged tracks having PID consistent with the pion
hypothesis that are not identified as electrons are considered
as pion candidates.

Neutral kaons are reconstructed via the decay $K_S^0\to\pi^+\pi^-$
with no PID requirements for these pions.
The two-pion invariant mass is required to be within 6~MeV$/c^2$
($\sim 2.5\sigma$) of the nominal $K^0$ mass and the displacement of
the $\pi^+\pi^-$ vertex from the IP in the transverse
($r-\phi$) plane is required to be between 0.2~cm and 20~cm. 
The direction from the IP to the $\pi^+\pi^-$ vertex is required to 
agree within 0.2 radians in the $r-\phi$ plane with the combined 
momentum of the two pions.
A pair of calorimeter showers not associated with charged tracks,
with an invariant mass within 15~MeV$/c^2$ of the nominal $\pi^0$ mass 
is considered as a $\pi^0$ candidate. An energy deposition of at least 
30~MeV and a photon-like shape are required for each shower.
$\kstar$ candidates are reconstructed from $\kpi$ pairs with an 
invariant mass within 50~MeV$/c^2$ of the nominal $\kstar$ mass.
We reconstruct $D^0$ mesons in the decay channels:
$\kpi$, $\kpipipi$ and $\kpipin$, using a requirement that the
invariant mass be within 20~MeV$/c^2$, 15~MeV$/c^2$ and 25~MeV$/c^2$
of the  nominal $D^0$ mass, respectively.
In each channel we further define a $D^0$ mass sideband region,
with width twice that of the signal region.
For the $\pi^0$ from the $\dkpipin$ decay, we require that 
its momentum in the CM frame be greater than 0.4~GeV$/c$ 
in order to reduce combinatorial background.
$D^{*0}$ mesons are reconstructed in the $\dsdpi$ decay mode. 
The mass difference between $D^{*0}$ and $D^0$ candidates is required 
to be within 4~MeV$/c^2$ of the expected value ($\sim 4\sigma$).

\begin{figure*}
  \includegraphics[width=0.45\textwidth] {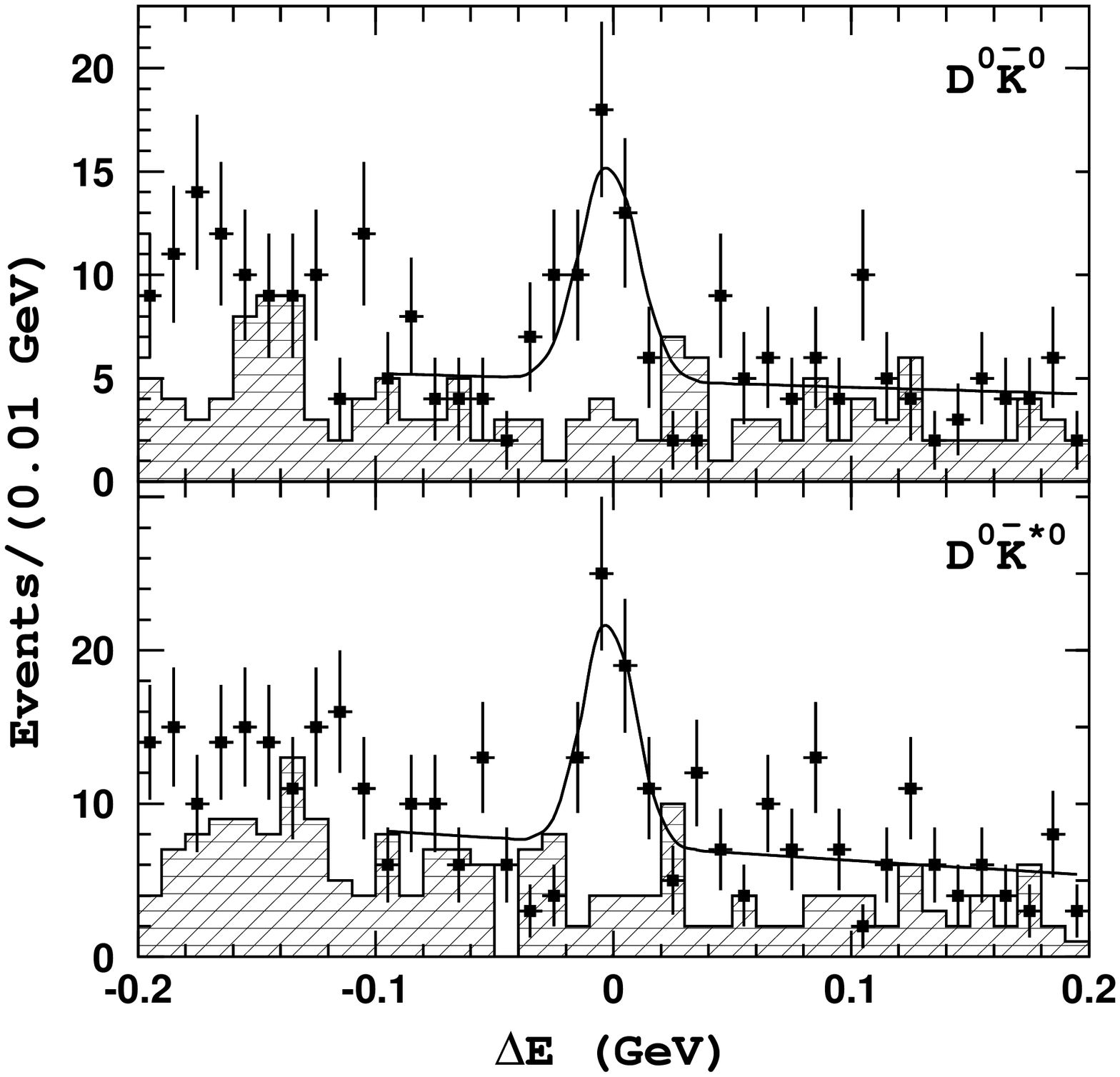} \hfill
  \includegraphics[width=0.45\textwidth] {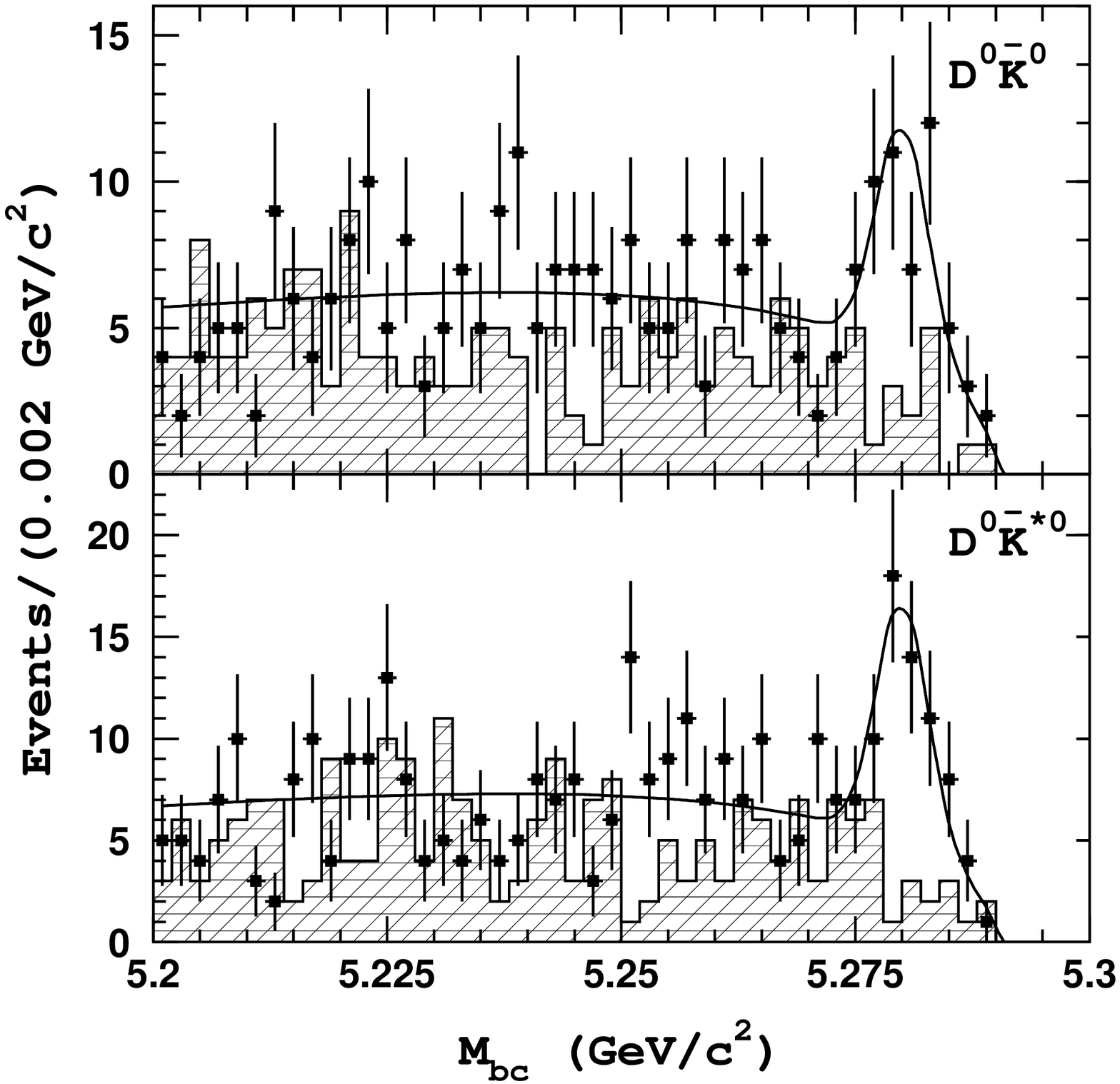} 
  \caption{$\de$ (left) and $\mbc$ (right) distributions for the 
    $\bdnkz$ candidates. Points with errors
    represent the experimental data, hatched histograms show
    the $D^0$ mass sidebands and curves are the results of the fits.}
  \label{mbc_all}
\end{figure*}

We combine $D^{(*)0}$ candidates with $K^0_S$ or $\kstar$ to form $B$ 
mesons. Candidate events are identified by their CM energy difference, 
$\de=(\sum_iE_i)-E_{\rm b}$, and the beam constrained mass, 
$\mbc=\sqrt{E_{\rm b}^2-(\sum_i\vec{p}_i)^2}$, where $E_{\rm b}$ is
the beam energy and $\vec{p}_i$ and $E_i$ are the momenta and energies 
of the $B$ meson decay products in the CM frame.
We select events with $\mbc>5.2$~GeV$/c^2$ and $|\de|<0.2$~GeV,
and define a $B$ signal region of
$5.272$~GeV$/c^2<\mbc<5.288$~GeV$/c^2$ and $|\de|<0.03$~GeV.
In the rare cases where there is more than one  candidate in an 
event, the candidate with the $D^{(*)0}$ and $\kz$ masses 
closest to their nominal values is chosen.
We use Monte Carlo (MC) simulation to model the response of
the detector and determine the efficiency~\cite{GEANT}.

To suppress the large combinatorial background dominated by 
the two-jet-like $e^+e^-\to\qq$ continuum 
process, variables that characterize the event topology are used. 
We require $|\cos\theta_{\rm thr}|<0.80$, where $\theta_{\rm thr}$ is 
the angle between the thrust axis of the $B$ candidate and that of the 
rest of the event.  This requirement eliminates 77\% of the continuum 
background and retains 78\% of the signal events. We also construct a 
Fisher discriminant, ${\cal F}$, which is based on the production
angle of the $B$ candidate, the angle of the $B$ candidate thrust axis 
with respect to the beam axis, and nine parameters that characterize 
the momentum flow in the event relative to the $B$ candidate thrust 
axis in the CM frame~\cite{VCal}. We impose a requirement on 
${\cal{F}}$ that rejects 67\% of the remaining continuum background 
and retains 83\% of the signal.

Among other $B$ decays, the most serious background comes from 
$B^0\to D^-\pi^+$, $D^-\to \kz K^-$, $\kz K^-\pi^0$,
$\kz K^-\pi^-\pi^+$ and $B^0\to D^-K^+$, $D^-\to \kz\pi^-$, 
$\kz\pi^-\pi^0$, $\kz\pi^-\pi^-\pi^+$. 
These decays produce the same final state as the $\bdzkz$ signal,
and their product branching fractions are up to ten times higher than 
those expected for the signal.
To suppress this type of background, we exclude candidates if the 
invariant mass of the combinations listed above is consistent
with the $D^-$ hypothesis within 25~MeV/$c^2$ ($\sim 3\sigma$).
The $\bar{B}^0\to D^{*+}K^-$, $D^{*+}\to D^0\pi^+$ decay can also
produce the same final state as the $\bdnkstar$ decay. But this decay 
is kinematically separated from the signal; the invariant mass selection
criteria for $\kstar$ candidates completely eliminates this background.
Another potential $\bb$ background comes from the
$\bar{B}^0\to D^{(*)0}\rho^0$ 
decay channel~\cite{asish} with one pion from $\rho^0$ decay
misidentified as a kaon. The reconstructed $\kpi$ invariant mass 
spectra for these events overlap with the signal $\kstar$ mass region, 
while their $\de$ distribution is shifted by about 70~MeV$/c^2$.
We study this background using MC simulation. The contribution to the 
$\bdzkstar$ signal region is found to be less than 0.2 events.
We examined the possibility that other $B$ meson decay modes might
produce backgrounds that peak in the signal region by studying a MC
sample of generic $\bb$ events that corresponds to about 1.5 times the 
data statistics. No other peaking backgrounds were found.


\begin{figure*}
  \includegraphics[width=0.245\textwidth] {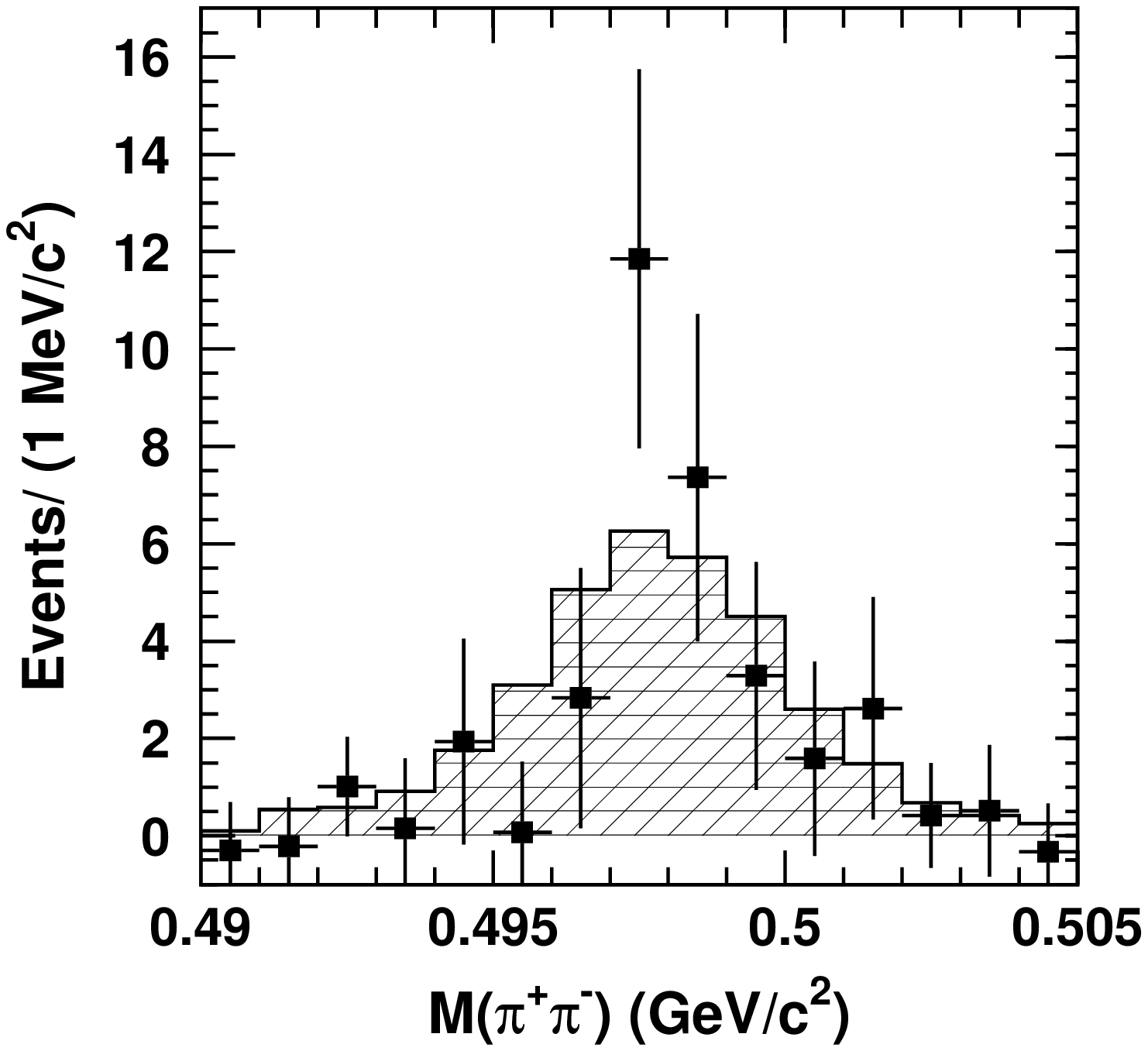} 
  \includegraphics[width=0.245\textwidth] {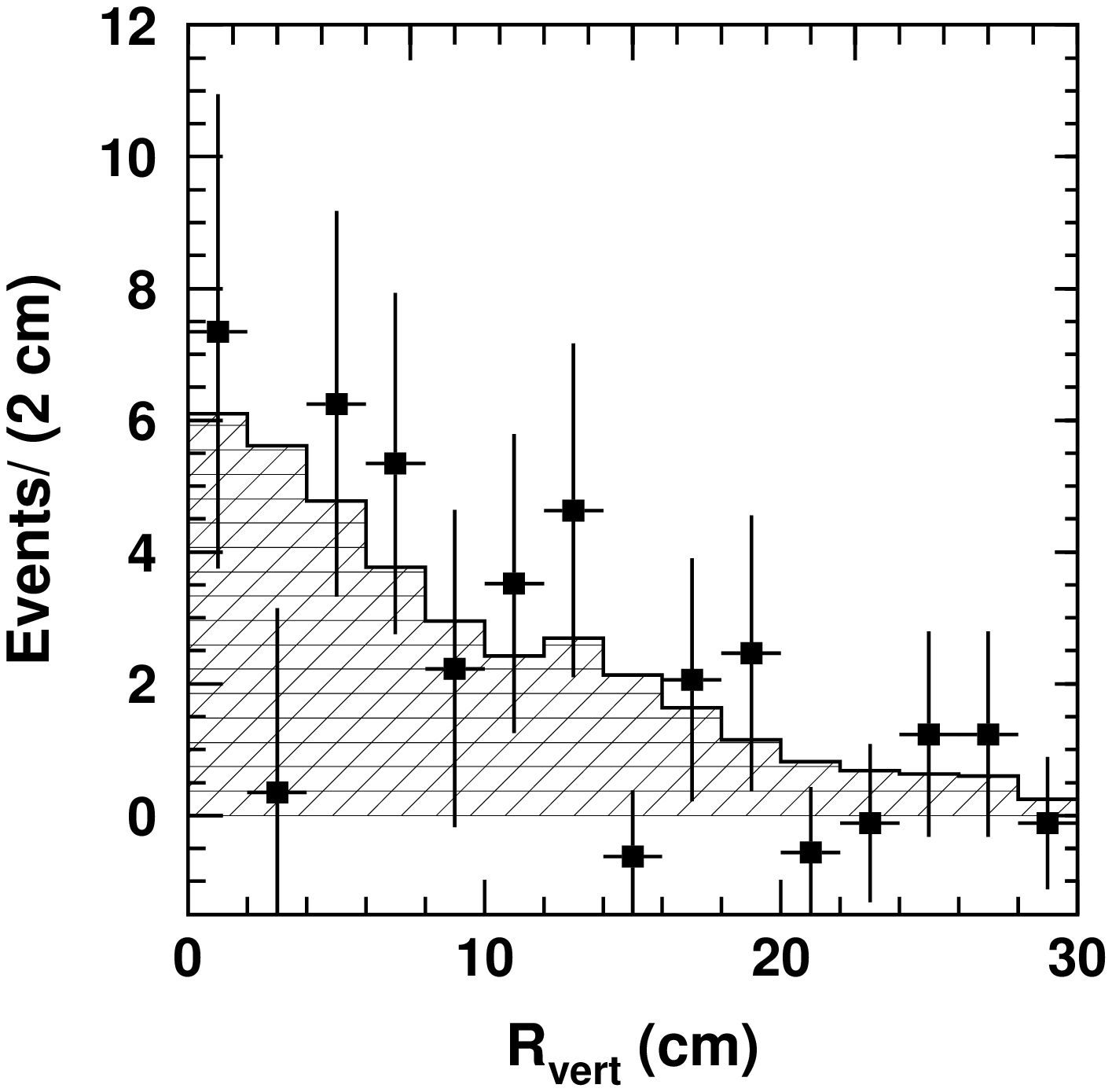}
  \includegraphics[width=0.245\textwidth] {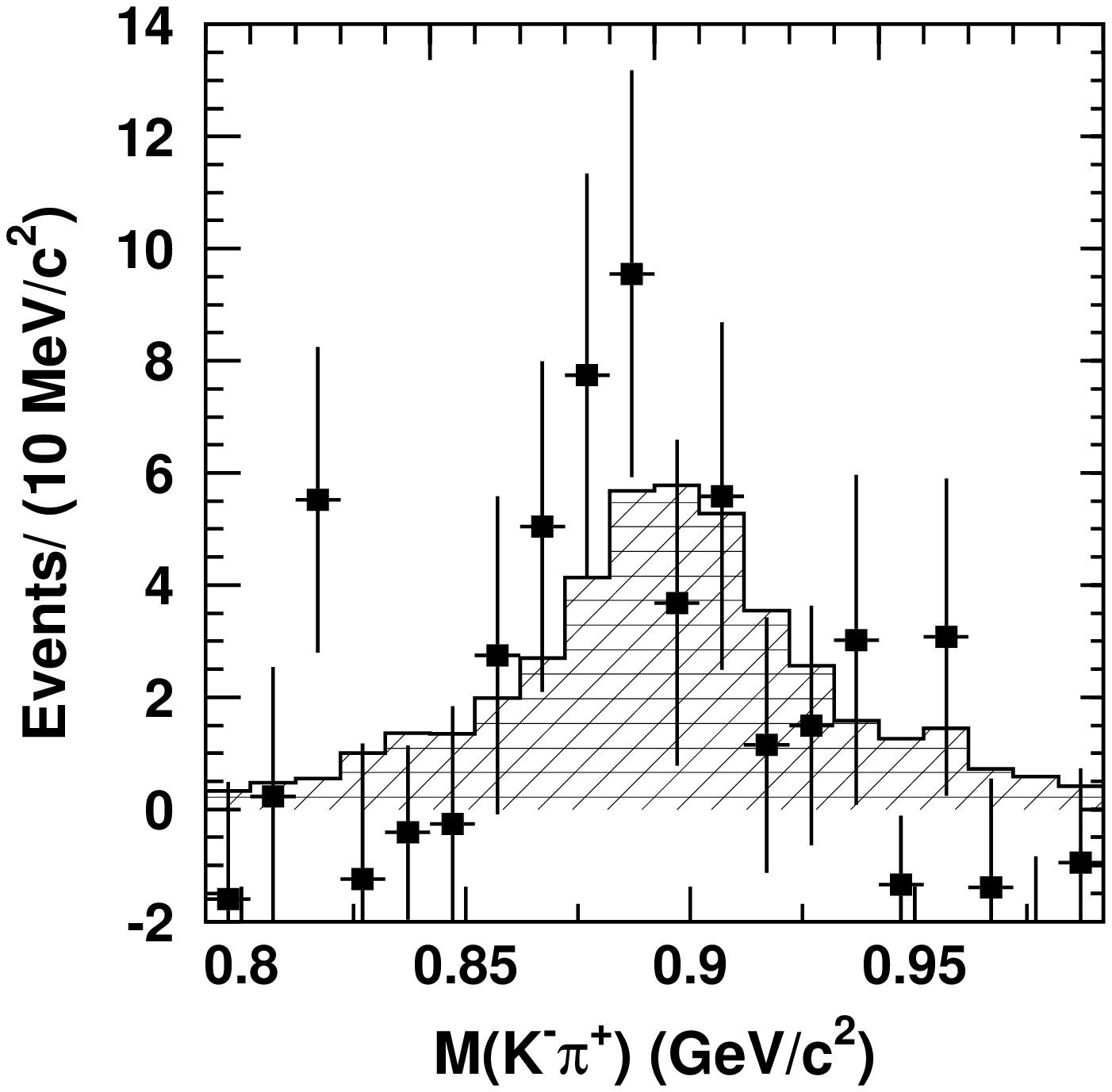} 
  \includegraphics[width=0.245\textwidth] {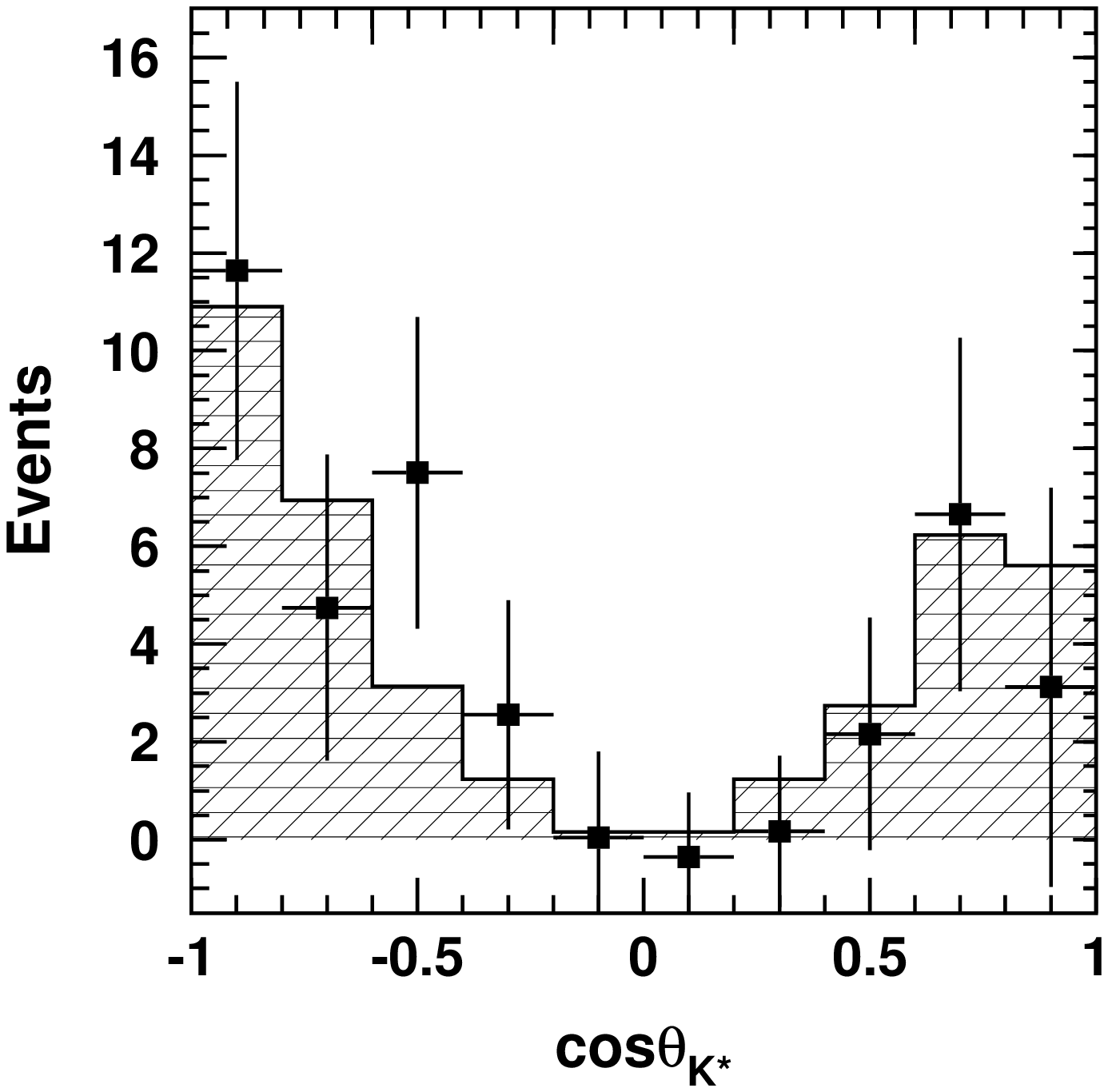}
  \caption{From left to right: $K^0_S$ candidates' invariant mass and
    flight distance for the $\bdnks$ channel, $\kstar$ candidates' 
    invariant mass and helicity distributions for the $\bdnkstar$ channel.
}
  \label{check}
\end{figure*}

The $\de$ and $\mbc$ distributions for $\bdnkz$ candidates are
presented in Fig.~\ref{mbc_all}, where all three $D^0$ decay modes are 
combined. Also shown in Fig.~\ref{mbc_all} by hatched histograms are
the distributions for events in $D^0$ mass sideband.
The sideband shape replicates the background shape well, confirming
that the background is mainly combinatorial in nature. 
Clear signals are observed for the $D^0\ks$ and $D^0\kstar$ final states.
As an additional cross check, we also study the $K^0_S$ candidates' 
invariant mass and flight distance distributions and $\kstar$ 
candidates' invariant mass and helicity distributions for these decays.
The helicity angle $\theta_{K^*}$ is defined as the angle between the
$\kstar$ momentum in the $B$ meson rest frame and the $K^-$ momentum
in the $\kstar$ rest frame.
The distributions mentioned above are shown in Fig.~\ref{check}, where 
points with error bars are the results of fits to the $\de$ spectra for 
experimental events in the corresponding bin, and histograms
are signal MC. All distributions are consistent with the MC expectation.

For each $D^0$ decay mode, the $\de$ distribution is fitted with a 
Gaussian for signal and a linear function for background. The Gaussian 
mean value and width are fixed to the values from MC simulation of
the signal events. The region $\de<-0.1$~GeV is excluded from the fit 
to avoid contributions from other $B$ decays, such as 
$B\to D^{(*)0}\kz (\pi)$ 
where $(\pi)$ denotes a possible additional pion.
For the $\mbc$ distribution fit we use the sum of a signal Gaussian
and an empirical background function with a kinematic 
threshold~\cite{argus}, with a parameter fixed from the analysis of the 
off-resonance data.
For the calculation of branching fractions, we use the signal yields
determined from the fit to the $\de$ distribution. This minimizes a
possible bias from other $B$ meson decays, which tend to peak in $\mbc$
but not in $\de$.
The fit results are presented in Table~\ref{defit}, where the listed
efficiencies include intermediate branching fractions.  
The statistical significance of the signal quoted in Table~\ref{defit} 
is defined as
$\sqrt{-2\ln({\cal L}_0/{\cal L}_{max})}$, where ${\cal L}_{max}$ and
${\cal L}_0$ denote the maximum likelihood with the nominal signal
yield and the signal yield fixed at zero, respectively.

\begin{figure}
  \includegraphics[width=0.45\textwidth] {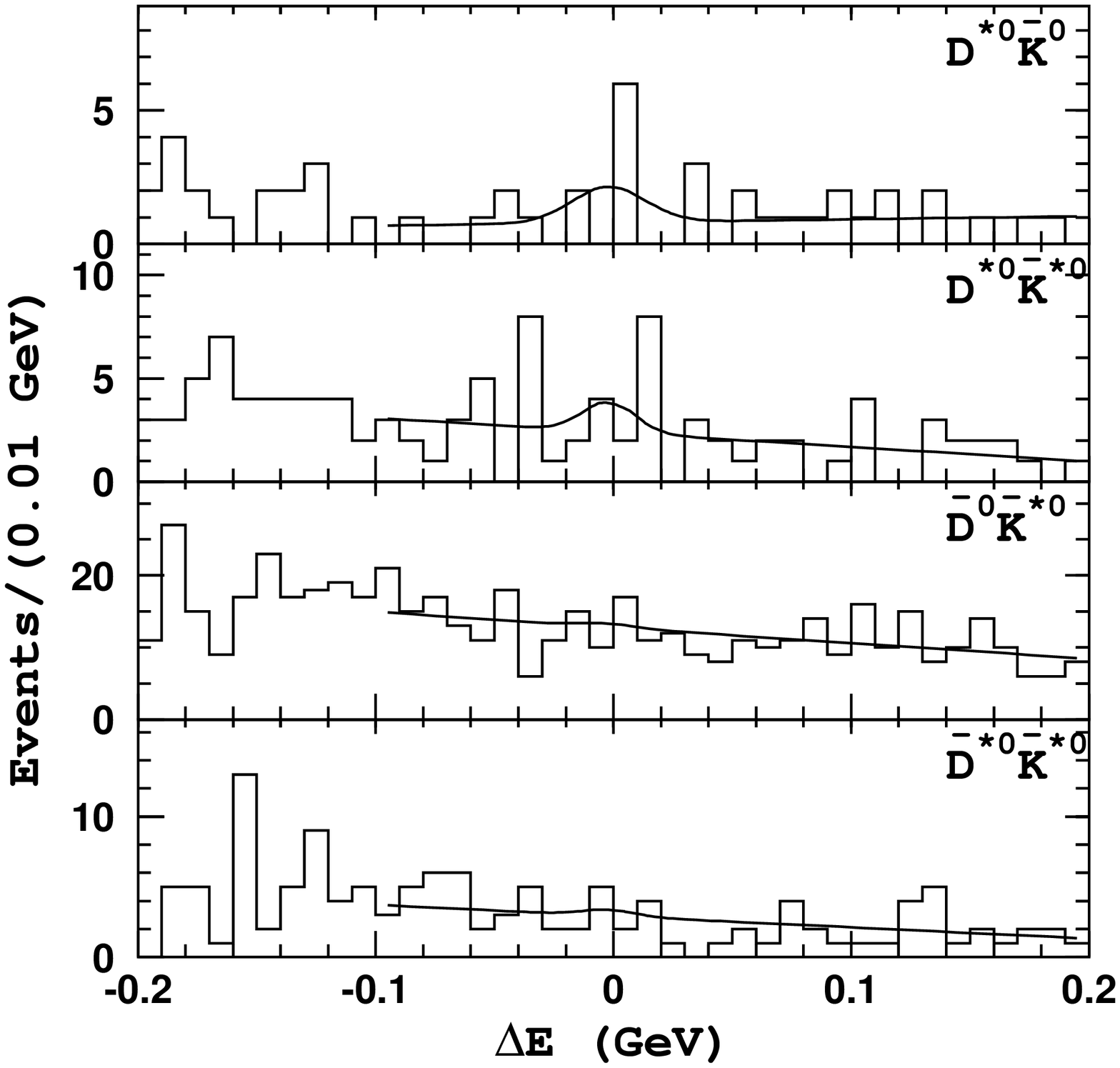}
  \caption{$\de$ distributions for the $\bdskz$ and $\bdzbarkstar$ 
    candidates. Open histograms represent the experimental data
    and curves show the results of the fits.}
  \label{de_ul}
\end{figure}

\begin{table*}
\caption{Fit results, efficiencies, branching fractions and
    statistical significances for $\bdzkz$ decays.}
\footnotesize
\medskip
\label{defit}
  \begin{tabular*}{\textwidth}{l@{\extracolsep{\fill}}ccccc}\hline\hline
 Mode  & $\de$ yield & $\mbc$ yield
& Efficiency ($10^{-3}$)  
& ${\cal B}$ ($10^{-5}$) & Significance\\\hline

$\bdnks$, $\dkpi$ & $9.3^{+4.2}_{-3.6}$ & $7.1^{+4.0}_{-3.4}$ & 
$2.50$ & $4.4^{+2.0}_{-1.7}\pm 0.5$ & $3.1\sigma$ \\
$\bdnks$, $\dkpipipi$ & $14.9^{+5.5}_{-4.9}$ & $13.1^{+5.2}_{-4.6}$ & 
$2.36$ & $7.4^{+2.7}_{-2.4}\pm 0.8$ & $3.6\sigma$ \\
$\bdnks$, $\dkpipin$ & $8.7^{+5.3}_{-4.8}$ & $7.0^{+4.2}_{-3.5}$ &
$2.52$ & $4.0^{+2.5}_{-2.2}\pm 0.4$ & $1.9\sigma$ \\\hline
$\bdnks$, simultaneous fit & $31.5^{+8.2}_{-7.6}$ & 
$27.0^{+7.6}_{-6.9}$ & $7.38$ & $5.0^{+1.3}_{-1.2}\pm 0.6$ & 
$5.1\sigma$ \\\hline

$\bdnkstar$, $\dkpi$ & $14.8^{+4.8}_{-4.1}$ & $11.7^{+4.6}_{-3.9}$ &
$3.47$ & $5.0^{+1.6}_{-1.4}\pm 0.6$ & $4.3\sigma$ \\
$\bdnkstar$, $\dkpipipi$ & $15.1^{+5.6}_{-5.0}$ & $13.4^{+5.4}_{-4.8}$ &
$3.34$ & $5.3^{+2.0}_{-1.8}\pm 0.6$ & $3.6\sigma$ \\
$\bdnkstar$, $\dkpipin$ & $9.9^{+6.4}_{-5.9}$ & $16.7^{+5.5}_{-4.9}$ &
$3.34$ & $3.5^{+2.3}_{-2.1}\pm 0.4$ & $1.7\sigma$ \\\hline
$\bdnkstar$, simultaneous fit & $41.2^{+9.0}_{-8.5}$ & 
$41.0^{+8.7}_{-8.1}$ & $10.15$ & $4.8^{+1.1}_{-1.0}\pm 0.5$ & 
$5.6\sigma$\\\hline

$\bdsks$, simultaneous fit & $4.2^{+3.7}_{-3.0}$ & $2.7^{+3.0}_{-2.4}$ &
$1.98$ & $<6.6$ 90\% CL & $1.4\sigma$ \\\hline

$\bdskstar$, simultaneous fit & $6.1^{+5.2}_{-4.5}$ & 
$8.6^{+4.2}_{-3.6}$ & $2.68$ & $<6.9$ 90\% CL & $1.4\sigma$ \\\hline

$\bdbarkstar$, simultaneous fit & $1.4^{+8.2}_{-7.6}$ & 
$9.2^{+7.7}_{-7.2}$ & $10.15$ & $<1.8$ 90\% CL & --- \\\hline

$\bdsbarkstar$, simultaneous fit & $1.2^{+4.1}_{-3.6}$ & 
$0.0^{+3.9}_{-3.2}$ & $2.68$ & $<4.0$ 90\% CL & --- \\\hline\hline

  \end{tabular*}
\end{table*}

For the final result we use a simultaneous fit to the $\de$ distributions 
for the three $D^0$ decay channels taking into account the corresponding 
detection efficiencies.
The normalization of the background in each $D^0$ sub-mode is
allowed to float while the signal yields are required to satisfy the 
constraint
$N_i= N_{\bb}\cdot{\cal B}(\bdzkz)\cdot\varepsilon_i\, ,$
where the branching fraction ${\cal B}(\bdzkz)$ is a fit parameter;
$N_{\bb}$ is the number of $\bb$ pairs
and $\varepsilon_i$ is the efficiency, which includes all
intermediate branching fractions.

The statistical significances for the $\bdnks$ and $\bdnkstar$ signals
are higher than $5\sigma$. The signals in the $\bdskz$
channels are not significant and we set 90\% confidence level
(CL) upper limits for these final states. 
We do not observe a significant signal for the $\bdzbarkstar$ decays and 
also present upper limits for them. 
Figure~\ref{de_ul} shows the $\de$ distributions
for $\bdskz$ and $\bdzbarkstar$ candidates.
The upper limit $N$ is calculated from the relation 
$\int^N_0 {\cal L}(n) dn=0.9\int^{\infty}_0 {\cal L}(n) dn$, 
where ${\cal L}(n)$ is the maximum likelihood with the signal 
yield equal to $n$. We take into account the systematic uncertainties
in these calculations by reducing the detection efficiency by one
standard deviation. 


As a check, we apply the same analysis procedure to 
$\bar{B}^0\to D^+[K_S^0 K^+]\pi^-$ and 
$\bar{B}^0\to D^{*+}[D^0\pi^+]K^-$ decay chains.
The estimated branching fractions of
$\br(\bdppi)=(2.5\pm 0.3)\times 10^{-3}$
and
$\br(\bar{B}^0\to D^{*+} K^-)=(1.7\pm 0.2)\times 10^{-4}$
(statistical errors only) ,
are consistent with the world average values~\cite{PDG}.

The following sources of systematic errors are found to be significant:
tracking efficiency (2\% per track), kaon identification
efficiency (2\%), $\pi^0$ efficiency (6\%), $K^0_S$ reconstruction 
efficiency (6\%), efficiency for slow pions 
from $\dsdpi$ decays (8\%), $D^{(*)0}$ branching fraction
uncertainties (2\% -- 6\%), signal and background shape
parameterization (4\%) and MC statistics (2\% -- 3\%).
The tracking efficiency error is estimated using
$\eta$ decays to $\gamma\gamma$ and $\pi^+\pi^-\pi^0$.
The kaon identification uncertainty is determined 
from $D^{*+}\to D^0\pi^+$, $D^0\to K^-\pi^+$ decays.
The $\pi^0$ reconstruction uncertainty is obtained using 
$D^0$ decays to $\kpi$ and $\kpipin$.
We assume equal production rates for $B^+B^-$ and $B^0\bar B^0$ pairs 
and do not include the uncertainty related to this assumption in the 
total systematic error. The overall systematic uncertainty is
found to be 11\% for $\bdnkz$ and 14\% for $\bdskz$.


In summary, we report the first observation of $\bdnkz$ decays. 
The branching fractions 
$\br (\bdnks)    =(5.0^{+1.3}_{-1.2}\pm 0.6)\times 10^{-5}$ and
$\br (\bdnkstar) =(4.8^{+1.1}_{-1.0}\pm 0.5)\times 10^{-5}$
are measured with $5.1\sigma$ and $5.6\sigma$ statistical significance,
respectively.
Note that we ignore the possible contribution of $\bar{B}^0\to D^0K^0$
to the former result, since we do not distinguish between 
$\ks$ and $K^0$.
No significant signal is observed in the $\bdskz$ final states. 
The corresponding upper limits at the 90\% CL are
${\cal{B}}(\bdsks)  <6.6\times 10^{-5}$ and
${\cal{B}}(\bdskstar) <6.9\times 10^{-5}$.
We also set the 90\% CL upper limits for the $V_{ub}$ suppressed
$\bdzbarkstar$ decays:
$\br(\bdbarkstar) <1.8\times 10^{-5}$ and
$\br(\bdsbarkstar) <4.0\times 10^{-5}$.

We wish to thank the KEKB accelerator group for the excellent
operation of the KEKB accelerator.
We acknowledge support from the Ministry of Education,
Culture, Sports, Science, and Technology of Japan
and the Japan Society for the Promotion of Science;
the Australian Research Council
and the Australian Department of Industry, Science and Resources;
the National Science Foundation of China under contract No.~10175071;
the Department of Science and Technology of India;
the BK21 program of the Ministry of Education of Korea
and the CHEP SRC program of the Korea Science and Engineering Foundation;
the Polish State Committee for Scientific Research
under contract No.~2P03B 17017;
the Ministry of Science and Technology of the Russian Federation;
the Ministry of Education, Science and Sport of the Republic of Slovenia;
the National Science Council and the Ministry of Education of Taiwan;
and the U.S. Department of Energy.


\begin{thebibliography}{99}
\bibitem{conj}{
The inclusion of charge conjugate states is implicit 
throughout this report. Note that we do not distinguish between final
states with $\ks$ and $K^0$.}
\end{thebibliography}
\end{document}